\documentclass[conference]{IEEEtran}
\makeatletter
\def\ps@headings{%
\def\@oddhead{\mbox{}\scriptsize\rightmark \hfil \thepage}%
\def\@evenhead{\scriptsize\thepage \hfil \leftmark\mbox{}}%
\def\@oddfoot{}%
\def\@evenfoot{}}
\makeatother
\newcommand{\nop}[1]{}
\newtheorem{example}{Example}

\usepackage{graphicx}
\usepackage{url}
\usepackage{float}
\usepackage{multirow}
\usepackage{amsmath}
\usepackage{subfigure}
\usepackage{balance}
\restylefloat{table}

\ifCLASSINFOpdf
\else
\fi
\begin{document}

%
\title{Battling the Internet Water Army: Detection of Hidden Paid Posters}

\author{\IEEEauthorblockN{Cheng Chen}
\IEEEauthorblockA{Dept. of Computer Science\\
University of Victoria\\
Victoria, BC, Canada}

\and
\IEEEauthorblockN{Kui Wu}
\IEEEauthorblockA{Dept. of Computer Science\\
University of Victoria\\
Victoria, BC, Canada}

\and
\IEEEauthorblockN{Venkatesh Srinivasan}
\IEEEauthorblockA{Dept. of Computer Science\\
University of Victoria\\
Victoria, BC, Canada}

\and
\IEEEauthorblockN{Xudong Zhang}
\IEEEauthorblockA{Dept. of Computer Science\\
Peking University \\
Beijing, China}

}

\maketitle

\begin{abstract}
We initiate a systematic study to help distinguish a special group of online users, called hidden paid posters, or termed ``Internet water army" in China, from the legitimate ones. On the Internet, the paid posters represent a new type of online job opportunity. They get paid for posting comments and new threads or articles on different online communities and websites for some hidden purposes, e.g., to influence the opinion of other people towards certain social events or business markets. Though an interesting strategy in business marketing, paid posters may create a significant negative effect on the online communities, since the information from paid posters is usually not trustworthy. When two competitive companies hire paid posters to post fake news or negative comments about each other, normal online users may feel overwhelmed and find it difficult to put any trust in the information they acquire from the Internet. In this paper, we thoroughly investigate the behavioral pattern of online paid posters based on real-world trace data. We design and validate a new detection mechanism, using both non-semantic analysis and semantic analysis, to identify potential online paid posters. Our test results with real-world datasets show a very promising performance.  
\end{abstract}

\begin{IEEEkeywords} Online Paid Posters, Behavioral Patterns, Detection 
\end{IEEEkeywords}



%
\IEEEpeerreviewmaketitle

\section{Introduction}


According to China Internet Network Information Center (CNNIC)~\cite{CNNIC}, there are currently around $457$ million Internet users in China, which is approximately $35\%$ of its total population. In addition, the number of active websites in China is over $1.91$ million. The unprecedented development of the Internet in China has encouraged people and companies to take advantage of the unique opportunities it offers. One core issue is how to make use of the huge online human resource to make the information diffusion process more efficient. Among the many approaches to e-marketing~\cite{chaffey2006internet}, we focus on {\em online paid posters} used extensively in practice.  

Working as an online paid poster is a rapidly growing job opportunity for many online users, mainly college students and the unemployed people. These paid posters are referred to as the ``Internet water army" in China because of the large number of people who are well organized to ``flood" the Internet with purposeful comments and articles. This new type of occupation originates from Internet marketing, and it has become popular with the fast expansion of the Internet. Often hired by public relationship (PR) companies, online paid posters earn money by posting comments and articles on different online communities and websites. Companies are always interested in effective strategies to attract public attention towards their products. The idea of online paid posters is similar to word-of-mouth advertisement. If a company hires enough online users, it would be able to create hot and trending topics designed to gain popularity. Furthermore, the articles or comments from a group of paid posters are also likely to capture the attention of common users and influence their decision. In this way, online paid posters present a powerful and efficient strategy for companies. To give one example, before a new TV show is broadcast, the host company might hire paid posters to initiate many discussions on the actors or actresses of the show. The content could be either positive or negative, since the main goal is to attract attention and trigger curiosity.

We would like to remark here that the use of paid posters extends well beyond China. According to a recent news report in the Guardian~\cite{usoperation}, the US military and a private corporation are developing a specific software that can be used to post information on social media websites using fake online identifications. The objective is to speed up the distribution of pro-American propaganda. We believe that it would encourage other companies and organizations to take the same strategy to disseminate information on the Internet, leading to a serious problem of spamming. 

However, the consequences of using online paid posters are yet to be seriously investigated. While online paid posters can be used as an efficient business strategy in marketing, they can also act in some malicious ways. Since the laws and supervision mechanisms for Internet marketing are still not mature in many countries, it is possible to spread wrong, negative information about competitors without any penalties. For example, two competitive companies or campaigning parties might hire paid posters to post fake, negative news or information about each other. Obviously, ordinary online users may be misled, and it is painful for the website administrators to differentiate paid posters from the legitimate ones. Hence, it is necessary to design schemes to help normal users, administrators, or even law enforcers quickly identify potential paid posters. 

Despite the broad use of paid posters and the damage they have already caused, it is unfortunate that there is currently no systematic study to solve the problem. This is largely because online paid posters mostly work ``underground'' and no public data is available to study their behavior. Our paper is the first work that tackles the challenges of detecting potential paid posters. We make the following contributions. 
\begin{enumerate}
	\item By working as a paid poster and following the instructions given from the hiring company, we identify and confirm the organizational structure of online paid posters similar to what has been disclosed before~\cite{4}.
	\item We collect real-world data from popular websites regarding a famous social event, in which we believe there are potentially many hidden online paid posters. 
	\item We statistically analyze the behavioral patterns of potential online paid posters and identify several key features that are useful in their detection.
	\item We integrate semantic analysis with the behavioral patterns of potential online paid posters to further improve the accuracy of our detection.
\end{enumerate}

\nop{Since online paid posters are supposed to post articles on different websites or post comments on popular news websites, our research will be based on real-world trace datasets, regarding articles and comments respectively. After collecting the required information, we define several parameters used to characterize their activities. Next, we have different analysis approaches for the two datasets in order to disclose hidden patterns of online paid posters.}

The rest of the paper is organized as follows. We present more background information and identify the organizational structure of online paid posters in Section~\ref{sec:background}. Section~\ref{sec:dataCollection} presents our data collection method. In Section~\ref{sec:analysis}, we statistically analyze non-semantic behavioral features of online paid posters. In Section~\ref{sec:SA}, we introduce a simple method for semantic analysis that can greatly help the detection of online paid posters. In Section~\ref{sec:detection}, we introduce our detection method and evaluation results. Related work is discussed in Section~\ref{sec:relatedwork}. We conclude the paper in Section~\ref{sec:conclusion}. 

\section{How Do Online Paid Posters Work?} \label{sec:background}

\subsection{Typical Cases}
To better understand the behavior and the social impact of online paid posters, we investigated several social events, which are likely to be boosted by online paid posters. We introduce two typical cases to illustrate how online paid posters could be an effective marketing strategy, in either a positive or a negative manner. 
\begin{example}
On July 16, 2009, someone posted a thread with blank content and a title of ``\textit{Junpeng Jia, your mother asked you to go back home for dinner!}'' on a Baidu Post Community of World of Warcraft, a Chinese online community for a computer game~\cite{2}. In the following two days, this thread magically received up to $300,621$ replies and more than $7$ million clicks. Nobody knew why this meaningless thread would get so much attention. Several days later, a PR company in Beijing claimed that they were the people who designed the whole event, with an intention to maintain the popularity of this online computer game during its temporary system maintenance. They employed more than $800$ online paid posters using nearly $20,000$ different user IDs. In the end, they achieved their goal-- even if the online game was not temporarily available, the website remained popular during that time and it encouraged more normal users to join. This case not only shows the existence of online paid posters, but also reveals the efficiency and effectiveness of such an online activity.
\end{example}
\begin{example}
On July 17, 2009, a Chinese IT company \textit{Qihu 360}, also known as \textit{360} for short, released a free anti-virus software and claimed that they would provide permanent anti-virus service for free. This immediately made \textit{360} a super star in anti-virus software market in China. Nevertheless, on July 29 an article titled ``\textit{Confessions from a retired employee of 360}" appeared in different websites. This article revealed some inside information about \textit{360} and claimed that this company was secretly collecting users' private data. The links to this post on different websites quickly attracted hundreds of thousands of views and replies. Though \textit{360} claimed that this article was fabricated by its competitors, it was sufficient to raise serious concerns about the privacy of normal users. Even worse, in late October, similar articles became popular again in several online communities. \textit{360} wondered how the articles could be spread so quickly to hundreds of online forums in a few days. It was also incredible that all these articles attracted a huge amount of replies in such a short time period. 

In 2010, \textit{360} and Tencent, two main IT companies in China, were involved in a bigger conflict. On September 27, \textit{360} claimed that Tencent secretly scans user's hard disk when its instant message client, QQ, is used. It thus released a user privacy protector that could be used to detect hidden operations of other software installed on the computer, especially QQ. In response, Tencent decided that users could no longer use their service if the computer had \textit{360}'s software installed. This event led to great controversy among the hundreds of thousands of the Internet users. They posted their comments on all kinds of online communities and news websites. Although both \textit{360} and Tencent claimed that they did not hire online paid posters, we now have strong evidence suggesting the opposite. Some special patterns are definitely unusual, e.g., many negative comments or replies came from newly registered user IDs but these user IDs were seldom used afterwards. This clearly indicates the use of online paid posters.  

Since a large amount of comments/articles regarding this conflict is still available in different popular websites, we in this paper focus on this event as the case study. 

\end{example}

\subsection{Organizational Structure of Online Paid Posters}

\subsubsection{Basic Overview}

These days, some websites, such as \textit{shuijunwang.com~}\cite{3}, offer the Internet users the chance of becoming online paid posters. To better understand how online paid posters work, Cheng, one of the co-authors of this paper, registered on such a website and worked as a paid poster. We summarize his experience to illustrate the basic activities of an online paid poster. 

Once online users register on the website with their Internet banking accounts, they are provided with a mission list maintained by the webmaster. These missions include posting articles and video clips for ads, posting comments, carrying out Q\&A sessions, etc., over other popular websites. Normally, the video clips are pre-prepared and the instructions for writing the articles/comments are given. There are project managers and other staff members who are responsible for validating the accomplishment of each poster's mission. Paid posters are rewarded only after their assignments pass the validation. An assignment is considered a ``fail'' if, for example, the posted articles or contents are deleted by other websites' administrators. In addition, there are some regular rules for the paid posters. For example, articles should be posted at different forums or at different sections of the same forum; Comments should not be copied and pasted from other users' replies; The mission should be finished on time (normally within 3 hours), and so on.  

Although the mission publisher has regulations for paid posters, they may not strictly follow the rules while completing their assignments, since they are usually rewarded based on the number of posts. That is why we can find some special behavioral patterns of potential paid posters through statistical analysis.     

\subsubsection{Management of Paid Posters}

\begin{figure} 
\centering
\includegraphics[scale=0.40]{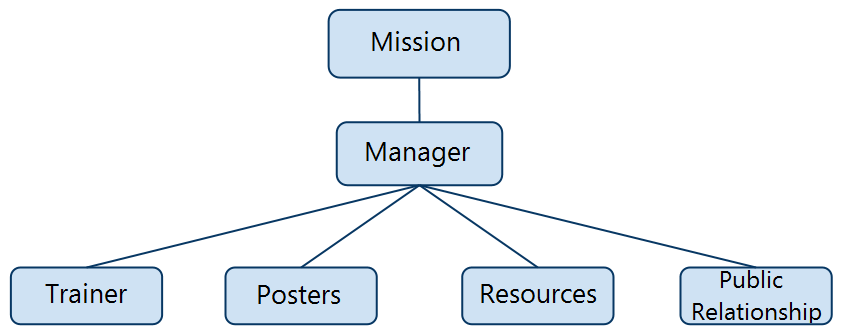}
\caption{Management structure of online paid posters} \label{fig:1}
\end{figure}

Occasionally, PR companies may hire many people and have a well-organized structure for some special events. Due to the large number of user IDs and different post missions, such an online activity needs to be well orchestrated to fulfill the goal. Our first-hand experience confirms an organizational structure of online paid posters as similar to that disclosed in~\cite{4}. When a mission is released, an organization structure as shown in Fig.~\ref{fig:1} is formed. The meaning or role of each component is as follows.

\begin{enumerate}
\item[-]\textit{\textbf{Mission}} represents a potential online event to be accomplished by online paid posters. Usually, $1$ project manager and $4$ teams, namely the trainer team, the poster team, the public relationship team, and the resource team are assigned to a mission. All of them are employed by PR companies. 

\item[-] \textit{\textbf{Project manager}} coordinates the activities of the four teams throughout the whole process.

\item[-] \textit{\textbf{Trainer team}} plans schedule for paid posters, such as when and where to post and the distribution of shared user IDs. Sometimes, they also accept feedback from paid posters.

\item[-] \textit{\textbf{Posters team}} includes those who are paid to post information. They are often college students and unemployed people. For each validated post, they get $30$ cents or $50$ cents. The posters can be grouped according to different target websites or online communities. They often have their own online communities for sharing experience and discussing missions.

\item[-] \textit{\textbf{Public relationship team}} is responsible for contacting and maintaining good relationship with other webmasters to prevent the posted messages from being deleted. Possibly, with some bonus incentives, these webmasters may even highlight the posts to attract more attention. In this sense, those webmasters are actually working for the PR companies.

\item[-] \textit{\textbf{Resources team}} is responsible for collecting/creating a large amount of online user IDs and other registration information used by the paid posters. Besides, they employ good writers to prepare specific post templates for posters.
\end{enumerate}


\section{Data Collection}\label{sec:dataCollection}

In this paper, we use the second example introduced in Section~\ref{sec:background}, the conflict between \textit{360} and Tencent, as the case study. We collected news reports and relevant comments regarding this special social event. While the number of websites hosting relevant content is large, most posts could be found at two famous Chinese news websites: Sina.com~\cite{6} and Sohu.com~\cite{sohu}, from which we collected enough data for our study. We call the data collected from Sina.com \textit{Sina dataset} and will use it as the training data for our detection model. The data collected from Sohu.com is called \textit{Sohu dataset} and it will be used as the test data for our detection method.

\nop{
As for online paid posters, posting articles and posting comments are two most common missions. Before doing research, we needed to figure out a specific social event. To this end, we chose to study the Case 2, \textit{‘A business conflict between Qi'hu \textit{360} and Tencent'}. Articles and comments information regarding this event will be collected and studied respectively. As mentioned before, these companies might hire paid posters to post negative information against each other.

\subsection{Articles dataset collected from Google}
For the articles posted by paid posters, one assumption is that an article would be repeatedly posted on different websites and communities with the same title, sometimes with different title. We can easily verify the result from the search engine. Firstly, we need to find some articles as our targets. Given carefully selected keywords that indicate attitude of pro or cons towards one of the company, we manually check the searching results to see if there are articles that appear more than 10 times on different websites. We found two popular articles which have this feature. Secondly, using the articles’ name, we collected URLs linked to the located websites, from the first $30$ pages of the searching results by Google. Each of the two articles has about $300$ records in total. 
 
In order to study the behavior of paid posters who post articles online, we manually check the detailed information of the posted articles via their links, including article’s title, poster’s ID, poster registration time, post time. These details will be used for our analysis in Section IV. Details have to be collected by hand because a web crawler seems useless when dealing with different web page formats.  
}

\nop{ collected from a famous Chinese news website using Gooseeker
For this part, data collection will be performed on a famous news websites, Sina.com\cite{6}. }

We searched all news reports and comments from Sina.com and Sohu.com over the time period from September 10, 2010 to November 21, 2010. As a result, we found $22$ news reports in Sina.com and $24$ news reports in Sohu.com. For each news report, there were many comments. For each comment, we recorded the following relevant information:  \textit{Report ID}, \textit{Sequence No.}, \textit{Post Time}, \textit{Post Location}, \textit{User ID}, \textit{Content}, and \textit{Response Indicator}, the meanings of which are explained in Table~\ref{tab:1}.

\begin{table}[!hbp]
\small
\centering
\caption{Recorded information for each comment} \label{tab:1}
{\renewcommand{\arraystretch}{1.5}
\begin{tabular}{|l|p{4.5cm}|}
\hline
Field & Meaning \\ \hline\hline
Report ID & The ID of news report that the comment belongs to\\
\hline
Sequence No. &  The order of the comment w.r.t. the corresponding news report \\ 
\hline
Post Time &  The time when the comment is posted\\
\hline 
Post Location & The location from where the comment was posted \\
\hline
User ID & The user ID used by the poster  \\
\hline
Content & The content of the comment \\
\hline
Response Indicator & Whether the comment is a new comment or a reply to another comment \\
\hline
\end{tabular}}
\end{table}

We were faced with several hurdles during the data collection phase. At the outset, we had to tackle the difficulty of collecting data from dynamic web pages. Due to the application of AJAX~\cite{webappwithAJAX} on most websites, comments are often displayed on web pages generated on the fly, and thus it was hard to retrieve the data from the source code of the web page. To be specific, after the client Internet explorer successfully downloads a HTML page, it needs to send further requests to the server to get the comments, which should be shown in the comment section. Most of the web crawlers that retrieve the source code do not support such a functionality to obtain the dynamically generated data. To avoid this problem, we adopted Gooseeker~\cite{5}, a powerful and easy-to-use software suitable for the above task. It allows us to indicate which part of the page should be stored in the disk and then it automatically goes through all the comment information page by page. In our case study, due to the popularity and the broad impact of this social event, some news reports ended up with more than $100$ pages of comments, with each page having $15$ to $20$ comments. We stored all the comments of one web page in a XML file. We then wrote a program in Python to parse all files to get rid of the HTML tags. We finally stored all the required information in the format described in Table~\ref{tab:1} into two separate files depending on whether the comments were from Sina.com or from Sohu.com. 

We then needed to clean up the data caused by some bugs on the server side of Sina and Sohu. We noticed that the server occasionally sent duplicate pages of comments, resulting in duplicate data in our final dataset. For example, for a certain report, we recorded more than $10,000$ comments, with nearly $5,000$ duplicate comments. After removing these duplicate data, we got $53,723$ records in Sina and $115,491$ records in Sohu. There was a special type of comments sent by mobile users with cellular phones. The user IDs of mobile users, no matter where they come from, are all labeled as ``Mobile User" on the web. There is no way to tell how many users are actually behind this unique user ID. For this reason, we have to remove all comments from ``Mobile User". We also needed to remove users who only posted very few comments, since it is hard to tell whether they are normal users or paid posters, even with manual check. To this end, we removed those users who only posted less than $4$ comments. Finally, Sohu allows anonymous posts (i.e., a user can post comments without needing to register for user ID). Since the real number of users behind the anonymous posts is unknown, we excluded these anonymous posts from our dataset.  

After the above steps, our Sina dataset included $552$ users and $20,738$ comments, and our Sohu dataset included $223$ users and $1,220$ comments. It is very interesting to see that the two datasets seem to have largely different statistical features, e.g., the average number of comments per user in the Sina dataset is about $37.6$ while that in the Sohu dataset is only $5.5$. One main reason is that Sohu allows anonymous posts, while Sina does not. 

A big question that we aim to answer: can we really build an effective detection system that is trained with one dataset and later works well for other datasets? We will disclose our findings in the following sections.

\section{Non-Semantic Analysis}\label{sec:analysis}
 
The goal of our non-semantic analysis is to find out the objective features that are useful in capturing potential paid posters' behavior. We use Sina dataset as our training data and thus we only perform statistical analysis on this dataset.  

First of all, we need to find the ground truth from the data: who are the paid posters? Based on our working experience as a paid poster, we manually selected $70$ ``potential paid posters" from the $552$ users, after reading the contents of their posts (many comments are meaningless or contradicting). We use the word \textit{potential} to avoid the non-technical argument about whether a manually selected paid poster is really a paid poster. Any absolute claim is not possible unless a paid poster admits to it or his employer discloses it, both of which are unlikely to happen. We stress that most detection mechanisms, such as email spam detection or forum spam detection~\cite{10}, face the same problem, and the argument whether an email should be really considered as a spam is usually beyond the technical scope.    

\nop{Then we manually select $70$ "potential paid posters" from the 552-user dataset. According to the experience working as a paid poster, these 70 users have the highest probability to be paid posters. We have to use the word "potential paid posters" because we currently do not have an efficient tool to verify whether they are or not. However, we can study the distribution of these detected users according to different parameters or features to find out the characteristics of behavior of the potential paid posters. We concluded some significant features of potential paid posters in the following. Meanwhile, there are 452 users we regard as normal users.}

After manually selecting the potential paid posters, we next perform statistical analysis to investigate objective features that are useful in capturing the potential paid posters' special behavior. We mainly test the following four features: percentage of replies, average interval time of posts, the number of days the user remains active and the number of news reports that the user comments on. In the following, we use $N_n$ and $N_p$ to denote the number of normal users and the number of potential paid posters who meet the test criterion, respectively. Additionally, we use $P_n$ and $P_p$ to denote the percentage of normal users and the percentage of potential paid posters who meet the test criterion, respectively.  

\nop{
    We consider these 70 selected abnormal users as our standard dataset of potential paid posters for the following statistical analysis. The analysis is performed in the following 5 aspects. For each of the 5 aspects, we firstly count the number of users who belong to a certain category, for instance, users who post comments in 2 different news reports. Then we figure out the number of abnormal users contained in each of the categories. We use N1 and N2 to represent the results while P1 and P2 are the percentages of them in 452-normal-user dataset and 70-abnormal-user dataset respectively.  }

\subsection{Percentage of Replies}

In this feature, we test whether a user tends to post new comments or reply to others' comments. We conjecture that potential paid posters may not have enough patience to read others' comments and reply. Therefore, they may create more new comments.  Table~\ref{table:replied} shows the statistical result and Fig.~ \ref{figure:1n} shows respective graphs, where $p$ represents the ratio of number of replies over the number of total comments from the same user.  

\begin{table}[H]
\small
\centering
\caption{The percentage of replies}
\begin{tabular}{|c|c|c|c|c|}
\hline
Criterion & $N_n$ & $P_n$& $N_p$ & $P_p$ \\\hline
\hline
$p <= 0.5$ & 121 & 26.77\% & 59 & 84.29\% \\ 
\hline
$p > 0.5$ & 331 & 73.23\% & 11 & 15.71\% \\  
\hline
\end{tabular}
\label{table:replied}
\end{table}

\begin{figure}[!ht]
\begin{center}
    \subfigure[The percentage of replies from normal users]{
    \includegraphics[width = 0.50\columnwidth]{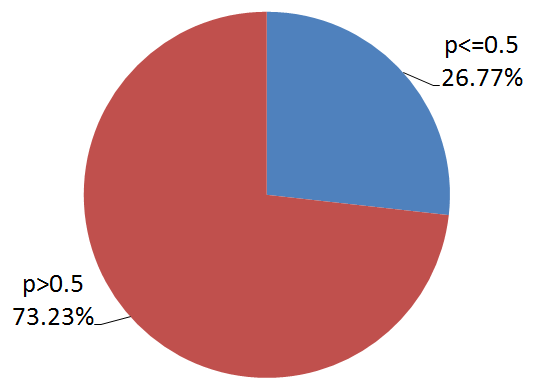}}
    \subfigure[The percentage of replies from potential paid posters]{
    \includegraphics[width = 0.45\columnwidth]{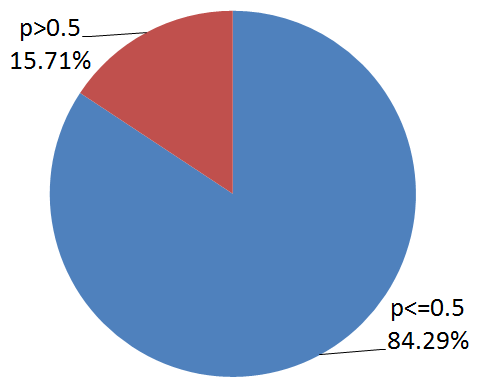}}
    \caption{The percentage of replies from normal users and potential paid posters}\label{figure:1n}
\end{center}
\end{figure}

\nop{

\begin{figure}[!hbp]
\centering
\includegraphics[scale=0.7]{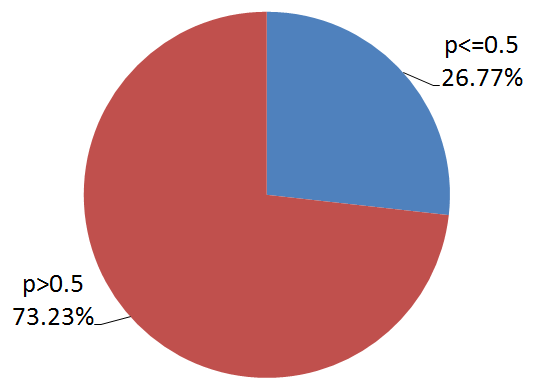}
\caption{normal users}
\label{figure:1n}
\end{figure}

\begin{figure}[!hbp]
\centering
\includegraphics[scale=0.7]{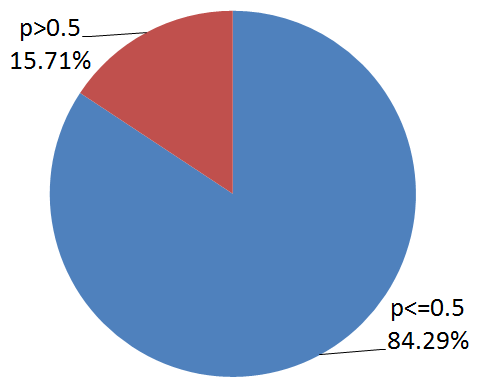}
\caption{potential paid posters}
\label{figure:1p}
\end{figure}
}

Based on the results, $59$ or $84.3\%$ potential paid posters have less than $50\%$ of posts being replies. In contrast, most normal users ($73.2\%$) posted more replies than new comments. This observation confirms our conjecture that potential paid posters are more likely to post new comments instead of reading and replying to others' comments. 

\subsection{Average Interval Time of Posts} 

We calculate the average interval time between two consecutive comments from the same user. Note that it is possible for a user to take a long break (e.g., several days) before posting messages again. To alleviate the impact of long break times, for each user, we divide his/her active online time into epochs. Within each epoch, the interval time between any two consecutive comments cannot be larger than $24$ hours. We calculate the average interval time of posts within each epoch, and then take the average again over all the epochs.  

Intuitively, normal users are considered to be less aggressive when posting comments while paid posters care more about finishing their jobs as soon as possible. This implies that the average interval time of posts from paid posters should be smaller. Table~\ref{table:inttime} shows the statistical results and Fig.~\ref{figure:2n} shows the corresponding graphs.

\begin{table}[H]
\small
\centering
\caption{The average interval time of posts}
\begin{tabular}{|c|c|c|c|c|}
\hline
Interval time (Second) & $N_n$ & $P_n$ & $N_p$ & $P_p$ \\ \hline\hline
0\url{~}150 & 103 & 22.91\% & 35 & 50.00\% \\\hline
150\url{~}300 & 153 & 33.94\% & 20 & 28.57\% \\\hline
300\url{~}450 & 93 & 20.67\% & 10 & 14.28\% \\\hline
450\url{~}600 & 41 & 9.16\% & 3 & 4.29\% \\\hline
600\url{~}750 & 35 & 7.83\% & 0 & 0.00\% \\\hline
750\url{~}900 & 11 & 2.52\% & 1 & 1.43\% \\\hline
$> 900$ & 13 & 2.97\% & 1 & 1.43\% \\\hline


\end{tabular}
\label{table:inttime}
\end{table}

\begin{figure}[!ht]
\begin{center}
    \subfigure[The average interval time of posts from normal users]{
    \includegraphics[width = 0.48\columnwidth]{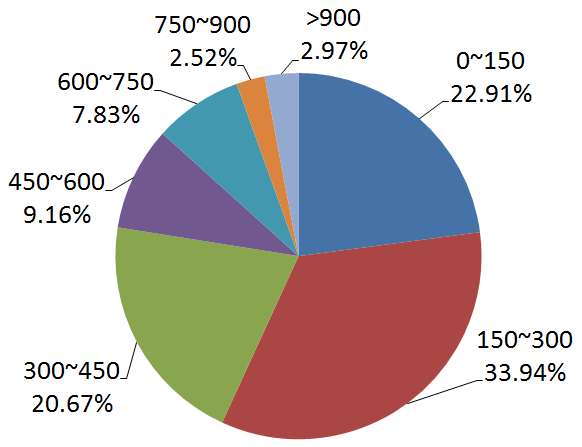}}
    \subfigure[The average interval time of posts from potential paid posters]{
    \includegraphics[width = 0.48\columnwidth]{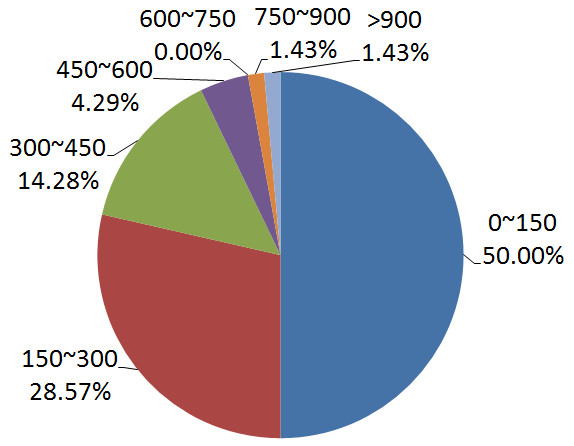}}
    \caption{The average interval time of posts from normal users and potential paid posters}\label{figure:2n}
\end{center}
\end{figure}

\nop{
\begin{figure}[!hbp]
\centering
\includegraphics[scale=0.6]{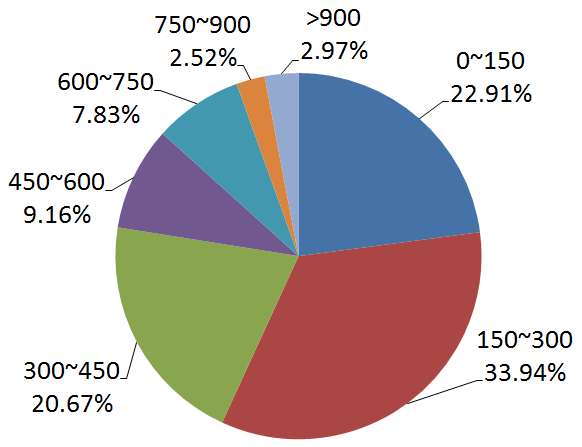}
\caption{normal users}
\label{figure:2n}
\end{figure}

\begin{figure}[!hbp]
\centering
\includegraphics[scale=0.6]{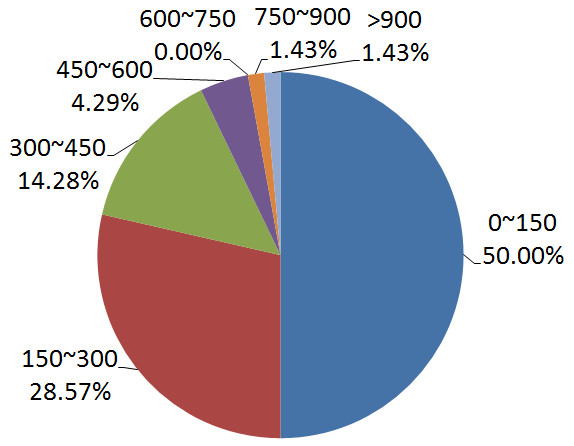}
\caption{potential paid posters}
\label{figure:2p}
\end{figure}
}

Based on the above result, $50\%$ of potential paid posters post comments with interval time less than $2.5$ minutes while $23\%$ of normal users post at such a speed. Nearly $80\%$ potential paid posters post comments with interval time less than $5$ minutes while only $57\%$ normal users post at this speed. From the figure, we can easily see that the potential paid posters are more likely to post in a very short time period. This matches our intuition that paid posters only care about finishing their jobs as soon as possible and do not have enough interest to get involved in the online discussion.

We observed that some potential paid posters also post messages in a relatively slow speed (the interval time is larger than $750$ seconds). There is one main explanation for the existence of these ``outliers". As mentioned earlier, the \textit{trainer team} may enforce rules that the paid posters need to follow. For example, identical replies should not appear more than twice in a same news report or within a short time period. Such rules are made to keep the paid posters from being detected easily. If a paid poster follows these tactics, he/she may have a statistical feature similar to that of a normal user. Nevertheless, it seems that the majority of potential paid posters did not follow the rules strictly.

\nop{For the interval time, another observation which is not reflected in the figures is that when users post their comments relatively slower, they also could be considered as paid posters because it would be one strategy to avoid being detected. As mentioned earlier, the mission scheduler will have different requirements for different missions. Specifically, for posting comments, the general rule is that each identical reply should not appear more than twice in a same news report and in a short period. However, paid posters sometimes do not follow this rule. If some users make a small change on their contents before posting, it makes the detection more easier because normal users, if they also want to post many comments to express their extreme feelings, won't have the intuition to make any change. Even though the time interval is larger, the posters who post similar comments are likely to be paid posters. In this case, the assumption is that paid posters would store their prepared comments and wait for a while for next posting to follow the rule, but normal users won't have to do like this. 
}

\subsection{Active Days} 

We analyzed the number of days that a user remains active online. We divided the users into $7$ groups based on whether they stayed online for $1, 2, 3, 4, 5, 6$ days and more than $6$ days, respectively. According to our experience as a paid poster, potential paid posters usually do not stay online using the same user ID for a long time. Once a mission is finished, a paid poster normally discards the user ID and never uses it again. When a new mission starts, a paid poster usually uses a different user ID, which may be newly created or assigned by the \textit{resource team}. Table~\ref{table:byday} shows the statistical result and Fig.~\ref{figure:3n} shows the corresponding graphs.

\begin{table}[H]
\small
\caption{The number of active days}
\centering
\begin{tabular}{|c|c|c|c|c|}
\hline
No. of active days & $N_n$ & $P_n$ & $N_p$ & $P_p$ \\\hline\hline
1 & 255 & 56.33\% & 43 & 61.43\% \\\hline 
2 & 99 & 21.91\% & 18 & 25.71\% \\\hline	
3 & 54 & 11.95\% & 8 & 11.43\% \\\hline 
4 & 28 & 6.19\% & 1 & 1.43\% \\\hline	
5 & 11 & 2.52\% & 0 & 0.00\% \\\hline	
6 & 3 & 0.66\% & 0 & 0.00\% \\\hline	
$>6$ & 2 & 0.44\% & 0 & 0.00\% \\\hline	
\end{tabular}

\label{table:byday}
\end{table}

\begin{figure}[!ht]
\begin{center}
    \subfigure[The number of active days of normal users]{
    \includegraphics[width = 0.48\columnwidth]{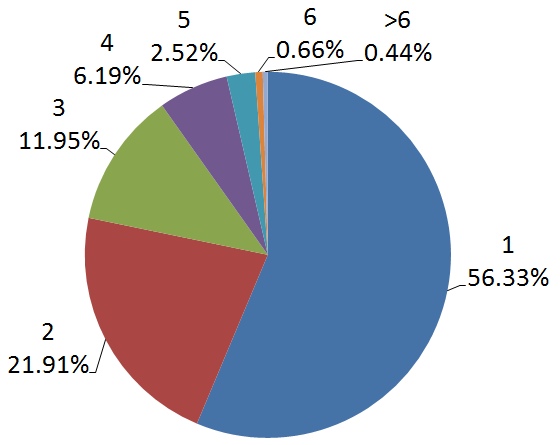}}
    \subfigure[The number of active days of potential paid posters]{
    \includegraphics[width = 0.48\columnwidth]{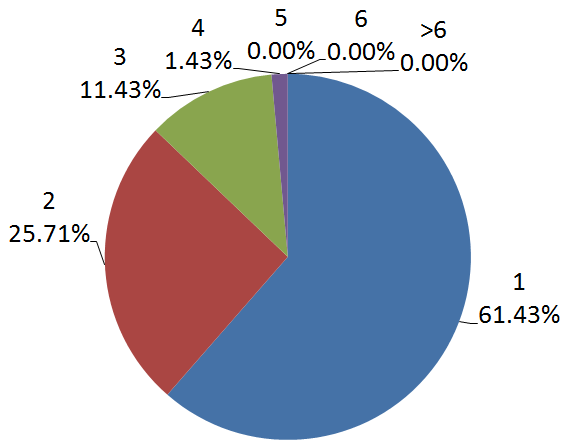}}
    \caption{The number of active days of normal users and potential paid posters}\label{figure:3n}
\end{center}
\end{figure}

\nop{

\begin{figure}[!hbp]
\centering
\includegraphics[scale=0.6]{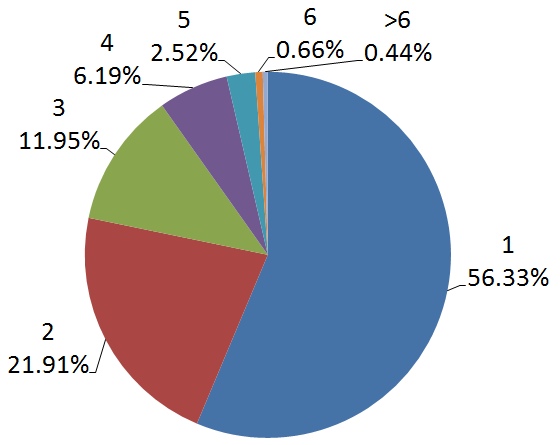}
\caption{normal users}
\label{figure:3n}
\end{figure}

\begin{figure}[!hbp]
\centering
\includegraphics[scale=0.6]{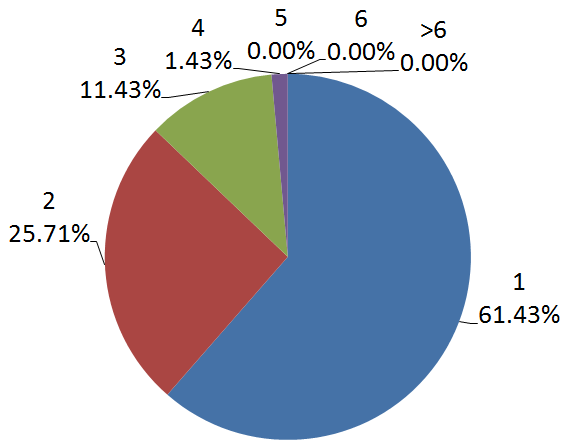}
\caption{potential paid posters}
\label{figure:3p}
\end{figure}
}

According to statistical result, the percentage of potential paid posters and the percentage of normal users are almost the same in the groups that remain active for $1, 2, 3,$ and $4$ days. Nevertheless, about $4\%$ of normal users keep taking part in the discussion for $5$ or more days, while we found no potential paid posters stayed for more than $4$ days. This evidence suggests that potential paid posters are not willing to stay for a long time. They instead tend to accomplish their assignments quickly and once it is done, they would not visit the same website again.  

\subsection{The Number of News Reports} 

We studied the number of news reports for which a user has posted comments. We divided the users into $7$ groups based on whether they have commented  on $1, 2, 3, 4, 5, 6$ or more news reports, respectively. Table~\ref{table:byreport} shows the statistical result and Fig.~\ref{figure:4n} shows the corresponding graphs.

\begin{table}[H]
\small
\caption{The number of news reports}
\centering
 
\begin{tabular}{|c|c|c|c|c|}
\hline
No. of News Reports & $N_n$ & $P_n$ & $N_p$ & $P_p$ \\\hline\hline
1 & 200 & 44.25\% & 31 & 44.29\% \\\hline 
2 & 114 & 25.22\% & 20 & 28.57\% \\\hline	
3 & 72 & 15.93\% & 9 & 12.85\% \\\hline 
4 & 39 & 8.63\% & 5 & 7.14\% \\\hline	
5 & 15 & 3.32\% & 3 & 4.29\% \\\hline	
6 & 8 & 1.77\% & 1 & 1.43\% \\\hline	
$>6$ & 4 & 0.88\% & 1 & 1.43\% \\\hline	
\end{tabular}

\label{table:byreport}
\end{table}

\begin{figure}[!ht]
\begin{center}
    \subfigure[The number of news reports that a normal user has commented]{
    \includegraphics[width = 0.48\columnwidth]{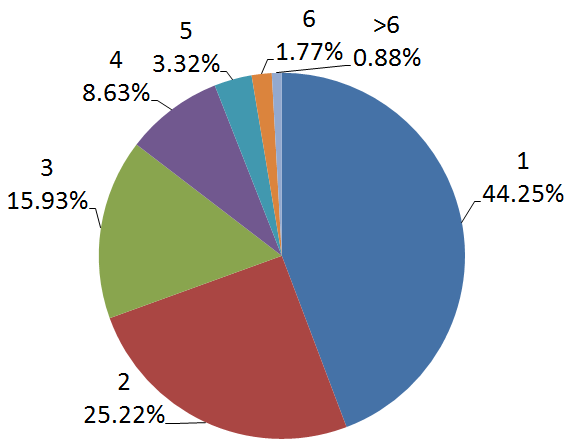}}
    \subfigure[The number of news reports that a potential paid poster has commented]{
    \includegraphics[width = 0.48\columnwidth]{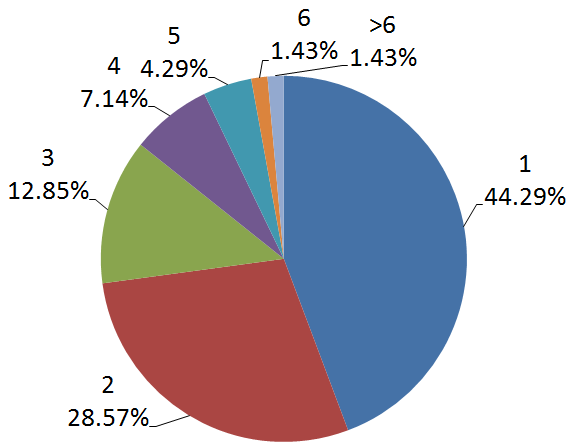}}
    \caption{The number of news reports that a user has commented}\label{figure:4n}
\end{center}
\end{figure}

\nop{
\begin{figure}[!hbp]
\centering
\includegraphics[scale=0.56]{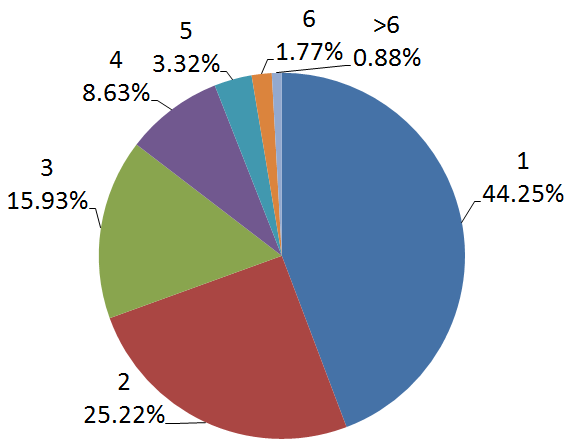}
\caption{normal users}
\label{figure:4n}
\end{figure}

\begin{figure}[!hbp]
\centering
\includegraphics[scale=0.56]{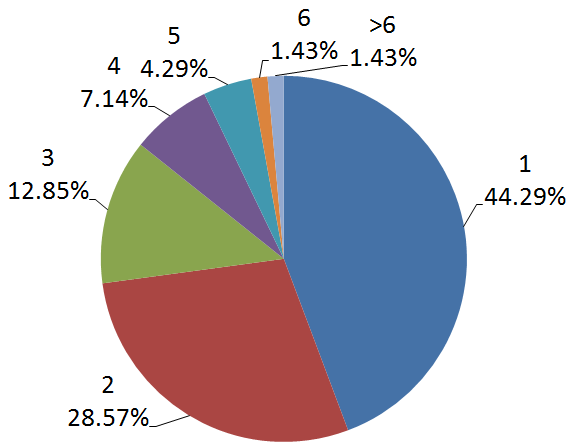}
\caption{potential paid posters}
\label{figure:4p}
\end{figure}
}

According to the result, the potential paid posters and normal users have similar distribution with respect to the number of commented news reports. We originally conjectured that paid posters might have a larger number of news reports that they post comments to. While normal users might not be interested in reports that are not well written or not interesting, paid posters care less about the contents of the news. Nevertheless, we did not find strong evidence to support this conjecture in the Sina dataset. This indicates that the number of commented news reports alone may not be a good feature for the detection of potential paid posters. 

\nop{
It looks like potential paid posters show a rather similar feature with normal users when it comes to the number of joined reports. It means that paid posters might click and read the articles just as what the normal users would do but paid posters would have different posting behaviors which we mentioned above. 
}

\subsection{Other Observations} 

We also discuss other possible features of potential paid posters. These observations come from our working experience as a paid poster. Although we cannot find sufficient evidence in the Sina dataset, we discuss these features as they can be beneficial for future research on this topic. 

First, there may be some pattern in geographic distribution of online paid posters. We performed statistical study on the Sina dataset, but found that both normal users and potential paid posters are mainly located in the center and the south regions of China. While the two companies involved in the event, Tencent and \textit{360}, are located in the province of Guang'dong and Beijing, respectively, we found no relationship between the locations of potential paid posters and the locations of the two companies. Fig.~\ref{figure:5n} shows the geographic distribution of normal users and potential paid posters, with the darker color representing more users. The figure does not exhibit a clear pattern to distinguish potential paid posters from normal users. 

\begin{figure}[!ht]
\begin{center}
    \subfigure[The geographical distribution of normal users]{
    \includegraphics[width = 0.46\columnwidth]{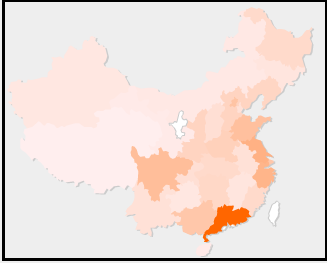}}
    \subfigure[The geographical distribution of potential paid posters]{
    \includegraphics[width = 0.46\columnwidth]{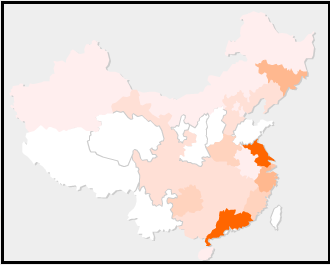}}
    \caption{The geographic distribution of users, with darker color representing more users}\label{figure:5n}
\end{center}
\end{figure}

\nop{ 
\begin{table}[!hbp]

\caption{statistic by location}
\centering

\begin{tabular}{|c|c|c|c|c|}
\hline
Area & n1 & p1 & n2 & p2 \\
\hline
guang dong & 75 & 0.166 & 11 & 0.157\\\hline 
jiang su & 34 & 0.075 & 11 & 0.157\\\hline	
zhe jiang & 32 & 0.071 & 5 & 0.071\\\hline	
si chuan & 27 & 0.060 & 2 & 0.029\\\hline	
shang hai & 22 & 0.049 & 4 & 0.057\\\hline	
shan dong & 26 & 0.058 & 0 & 0.000\\\hline	
bei jing & 24 & 0.053 & 1 & 0.014\\\hline	
guang xi & 22 & 0.049 & 2 & 0.029\\\hline	
he nan & 17 & 0.038 & 3 & 0.043\\\hline	
he bei & 15 & 0.033 & 2 & 0.029\\\hline 
liao ling & 14 & 0.031 & 2 & 0.029\\\hline 	
fu jian & 12 & 0.027 & 3 & 0.043\\\hline	
hu nan & 13 & 0.029 & 2 & 0.029\\\hline	
shan3xi & 15 & 0.033 & 0 & 0.000\\\hline	
hu bei & 12 & 0.027 & 2 & 0.029\\\hline	
hei long jiang & 12 & 0.027 & 1 & 0.014\\\hline 
chong qing & 9 & 0.020 & 2 & 0.029\\\hline 
ji lin & 6 & 0.013 & 5 & 0.071\\\hline 
an hui & 8 & 0.018 & 1 & 0.014\\\hline 
shan1xi & 9 & 0.020 & 0 & 0.000\\\hline 
tian jin & 8 & 0.018 & 1 & 0.014\\\hline 
jiang xi & 6 & 0.013 & 2 & 0.029\\\hline 
gui zhou & 5 & 0.011 & 3 & 0.043\\\hline 
yun nan & 8 & 0.018 & 0 & 0.000\\\hline 
xin jiang & 5 & 0.011 & 1 & 0.014\\\hline 
nei meng & 4 & 0.009 & 1 & 0.014\\\hline 
gan su & 3 & 0.007 & 2 & 0.029\\\hline	
hai nan & 3 & 0.007 & 1 & 0.014\\\hline	
USA & 2 & 0.004 & 0 & 0.000\\\hline	
Singapore & 1 & 0.002 & 0 & 0.000\\\hline 
qing hai & 1 & 0.002 & 0 & 0.000\\\hline 
Canada & 1 & 0.002 & 0 & 0.000\\\hline 
xi zang & 1 & 0.002 & 0 & 0.000\\ 

\hline

\end{tabular}
\label{table:location}
\end{table}

\begin{figure}[!hbp]
\centering
\includegraphics[scale=0.4]{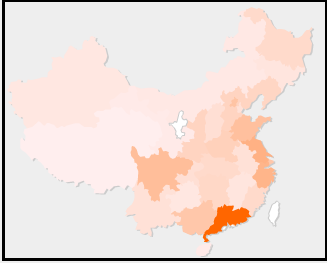}
\caption{normal users}
\label{figure:5n}
\end{figure}

\begin{figure}[!hbp]
\centering
\includegraphics[scale=0.4]{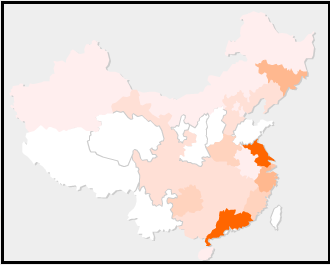}
\caption{potential paid posters}
\label{figure:5p}
\end{figure}

According to the table and figures, provinces of Guang'dong and Jiang'su have the majority of potential paid posters while Guang'dong has the most normal users. Unfortunately, we didn't see clear relationship between location and paid posters except Guang'dong and Jiang'su.  
}

Second, the same user ID appears at different geographical locations within a very short time period. This is a clear indication of paid poster. Normal users are not able to move to a different city in a few minutes or hours, but paid posters can because their user IDs may be assigned dynamically by the \textit{resource team}. We identified this possible feature for analysis but could not find sufficient evidence in the Sina dataset. 

Third, there might exist contradicting comments from paid posters. The reason is that they are paid to post without any personal emotion. It is their job. Sometimes, they just post comments without carefully checking their content. Nevertheless, this feature requires the detection system to have enough intelligence to understand the meaning of the comments. Incorporating this feature into the system is challenging.

Fourth, paid posters may post replies that have nothing to do with the original message. To earn more money, some paid posters just copy and paste existing posts and simply click the \textit{reply} button to increase the total number of posts. They do not really read the news reports or others' comments. Again, this feature is hard to implement since it requires high intelligence for the detection system.

\section{Semantic Analysis} \label{sec:SA}

An important criterion in our manual identification of a potential paid poster is to read his/her comments and make a choice based on common sense. For example, if a user posted meaningless messages or messages contradicting each other, the user is very likely to be a paid poster. Nevertheless, it is very hard to integrate such human intelligence into a detection system. In this section, we propose a simple semantic analysis method that is demonstrated to be very effective in detecting potential paid posters. 

While it is hard to design a detection system that understands the meaning of a comment, we observed that potential paid posters tend to post similar comments on the web. In many cases, a potential paid poster may copy and paste existing comments with slight changes. This provides the intuition for our semantic analysis technique. 

Our basic idea is to search for similarity between comments. To do this, we first need to overcome the special difficulty in splitting a Chinese sentence into words. Unlike English sentences that have a space between words, many languages in Asia such as Chinese and Japanese depend on context to determine words. They do not have space between words and how to split a sentence is left to the readers. We used a famous Chinese splitting software, called ICTCLAS2011~\cite{8}, to cut a sentence into words. For a given sentence, the software outputs its content words and stop words~\cite{stopwords}. Simply put, content words are words that have an independent meaning, such as noun, verb, or adjective. They have a stable lexical meaning and should express the main idea of a sentence. Stop words are words that do not have a specific meaning but have syntactic function in the sentence to make it grammatically correct. Stop words thus should be filtered out from further processing. 

The above step translates a sentence into a list of content words. For a given pair of comments, we compare the two lists of content words. As mentioned before, a paid poster may make slight changes before posting two similar comments. Therefore, we may not be able to find an exact match between the two lists. We first find their common content words, and if the ratio of the number of common content words over the length of the shorter content word list is above a threshold value (e.g., $80\%$ in our later test), we conclude that the two comments are similar. If a user has multiple pairs of similar comments, the user is considered a potential paid poster. Note that similarity of comments is not transitive in our method. 

We found that a normal user might occasionally have two \textit{identical} comments. This may be caused by the slow Internet access, due to which the user presses the \textit{submit} button twice before his/her post is displayed. Our manual check of these users confirmed that they are normal users, based on the content they posted. To reduce the impact of the ``unusual behavior of normal users", we set the threshold of similar pairs of comments to $3$. This threshold value is demonstrated to be effective in addressing the above problem. 

While there are many other complex semantic analysis methods to represent the similarity between two texts~\cite{li2006sentence}~\cite{higgins2007sentence}~\cite{achananuparp2008evaluation}~\cite{jing2008chinese}, we believe that comments are much shorter than articles and therefore a simple method as above would be good enough. This is demonstrated later in Section~\ref{sec:detection}.

\begin{figure}[!hbp]
\begin{center}
    \subfigure[The number of similar pairs of comments posted by normal users]{
    \includegraphics[width = 0.46\columnwidth]{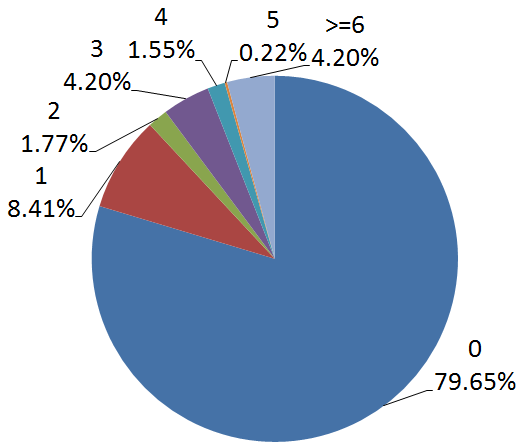}}
    \subfigure[The number of similar pairs of comments posted by potential paid posters]{
    \includegraphics[width = 0.46\columnwidth]{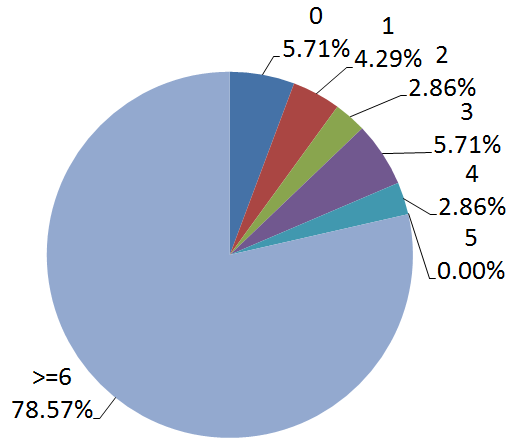}}
    \caption{The number of similar pairs of comments posted by a user}\label{figure:sa}
\end{center}
\end{figure}

We performed the semantic analysis over the training data, Sina dataset. The result is shown in Table~\ref{table:bysemantic} and Fig.~\ref{figure:sa} shows the corresponding graphs. We list the statistic result regarding the number of similar comment pairs of each user. From this table, we can see that normal users tend to post different comments. $79.6\%$ of them do not have any similar comment pairs. In sharp contrast, $78.6\%$ of the potential paid posters have more than $5$ similar comment pairs!

\begin{table}[H]
\caption{Semantic analysis of Sina dataset}
\small
\centering
\begin{tabular}{|c|c|c|c|c|}
\hline
Similar Pairs of Comments  & $N_n$ & $P_n$ & $N_p$& $P_p$ \\\hline\hline
0 & 360 & 79.65\% & 4 & 5.71\% \\\hline
1 & 38 & 8.41\% & 3 & 4.29\% \\\hline
2 & 8 & 1.77\% & 2 & 2.86\% \\\hline
3 & 19 & 4.20\% & 4 & 5.71\% \\\hline
4 & 7 & 1.55\% & 2 & 2.86\% \\\hline
5 & 1 & 0.22\% & 0 & 0.00\% \\\hline
$>=6$ & 19 & 4.20\% & 55 & 78.57\% \\\hline
\end{tabular}
\label{table:bysemantic}
\end{table}

\section{Classification}\label{sec:detection}

The objective of our classification system is to classify each user as a potential paid poster or a normal user using the features investigated in Section~\ref{sec:analysis} and Section~\ref{sec:SA}. According to the statistical and semantic analysis results, we found that any single feature is not sufficient to locate potential paid posters. Therefore, we use and compare the performance of different combinations of the five features discussed in the previous two sections in our classification system. We model the detection of potential paid posters as a binary classification problem and solve the problem using a support vector machine (SVM)~\cite{cristianini2006introduction}.

We used a Python interface of LIBSVM~\cite{7} as the tool for training and testing. By default, LIBSVM adopts radial basis function~\cite{cristianini2006introduction} and a $10$-fold cross-validation method to train the data and obtain a classifier. After training the classifier with the Sina dataset, we used the classifier to test the Sohu dataset. 

Before evaluating the performance of our classifier on the test dataset, we first manually identify the potential paid posters in the Sohu dataset, by reading the contents of their posts. The number of manually selected paid posters and normal users are listed in Table~\ref{table:sina_sohu}. 

\begin{table}[H]
\small
\centering

\begin{tabular}{|c|c|c|}

\hline

 & Sina.com & Sohu.com \\\hline

Paid Poster & 70 & 82 \\\hline 

Normal Users & 452 & 141 \\\hline	

\end{tabular}

\caption{Number of paid posters and normal users in Sina.com and Sohu.com}

\label{table:sina_sohu}

\end{table}
\vspace{-0.05in}

We evaluate the performance of the classifier using the four metrics: \textit{precision}, \textit{recall}, \textit{F-measure} and \textit{accuracy}, which are defined in Table~\ref{table:classifyMetrics}. Note that these four metrics are well known and broadly used measures in the evaluation of a classification system~\cite{sokolova2006beyond}. In the table, \textit{benchmark result} means the result obtained with manual identification of potential paid posters.  



\begin{table}[H]
\small
\centering


\renewcommand{\multirowsetup}{\centering} 

\begin{tabular}{|c|c|c|c|}

\hline

\multicolumn{2}{|c|}{} & \multicolumn{2}{|c|}{Classified Result} \\ \cline{3-4}

\multicolumn{2}{|c|}{} & Normal User & Paid Poster\\\hline

\multirow{2}{1.5cm}{Benchmark Result} & Normal User & True Negative & False Positive  \\\cline{2-4}

 & Paid Poster & False Negative & True Positive \\\hline


\end{tabular}

\begin{align}
Precision &= \frac{True Positive}{True Positive + False Positive}  \nonumber\\
Recall &= \frac{True Positive}{True Positive + False Negative} \nonumber\\
F-measure &= 2*\frac{Precision*Recall}{Precision + Recall}\nonumber \\
Accuracy &= \frac{True Negative + True Positive}{Total Number of Users}\nonumber
\end{align}

\caption{Metrics to evaluate the performance of a classification system}

\label{table:classifyMetrics}

\end{table}

\subsection{Classification without Semantic Analysis}
\nop{
The classification only examines the case of comment posters rather than article posters. In our view, the behavior of article posters are less complicated than comment posters. Meanwhile, we have more abundant dataset on comment posters which can be used to do detailed research. 
}

We firstly focus on the classification only using statistical analysis results. Based on the statistical analysis in Section~\ref{sec:analysis}, we notice that the first two features, ratio of replies and average interval time of posts, show great difference between the potential paid posters and the normal users. Therefore, we train the SVM model using the Sina dataset with those two features. We test the model with the Sohu dataset to see the performance. As a comparison, we also train the model using all the four non-semantic features. The results are listed in Table~\ref{table:classify1}.
\nop{  we don't use the location feature since it is hard to present the relationship between different places using numbers. }

\nop{For the datasets, we use the one from sina.com as training set while another one from sohu.com\cite{sohu} is taken as testing set. We perform the exactly same statistical analysis on Sohu dataset, including all the features listed in Section \ref{section:sa}. The number of manually picked out paid posters and normal users are listed in Table \ref{table:sina_sohu}. }

\nop{The test results of SVM classification on the Sohu dataset are listed in Table~\ref{table:classify1}.}

\begin{table}[H]
\small
\centering

\begin{tabular}{|c|c|c|c|}

\hline
Metrics & 2-Feature & 4-Feature & 5-Feature \\\hline\hline

True Negative & 141 & 108 & 138 \\\hline

False Positive & 0 & 33 & 3\\\hline

False Negative & 80 & 50 & 22\\\hline

True Positive & 2 & 32 & 60 \\\hline

Precision & 100.00\% & 49.23\% & 95.24\% \\\hline 

Recall & 2.43\% & 39.02\% & 73.17\% \\\hline	

F-measure & 4.76\% & 43.54\% & 82.76\% \\\hline

Accuracy: & 64.12\% & 62.78\% & 88.79\% \\\hline

\end{tabular}

\caption{Test results with non-semantic and semantic features}

\label{table:classify1}

\end{table}

For the 2-feature test, although the precision is $100\%$, only $2$ out of $82$ potential paid posters are correctly identified by the classifier. It will be unacceptable if we want to use this classifier to find out paid posters. These results suggest that the first two features lead to significant bias in our classification, and we need to add more features to our classifier. 

When we use the four non-semantic features as vectors to train the SVM model and do the same test on Sohu dataset, the results are much improved except the precision and the accuracy. Nevertheless, we can see that the values of false positive and false negative are too high to claim acceptable performance. The low precision result indicates that the SVM classifier using the four non-semantic features as its vector set is unreliable and needs to be improved further. We achieve this by adding the semantic analysis to our classifier.

\nop{

Notice that the statistical analysis in Section \ref{section:sa} don't have semantic analysis which means that we don't take advantage of the information included in the content of comments. We will firstly show the performance of SVM classification based on these four features. The predicting result from LIBSVM are listed in Table~\ref{table:classify2}.
}

\nop{
\begin{table}[H]

\centering

\begin{tabular}{|c|c|}

\hline

True Negative & 108 \\\hline

False Positive & 33 \\\hline

False Negative & 50 \\\hline

True Positive & 32 \\\hline

Precision & 49.23\% \\\hline 

Recall & 39.02\% \\\hline	

F-measure & 43.54\% \\\hline

Accuracy: & 62.78\% \\\hline

\end{tabular}

\caption{Test results with the four non-semantic features}

\label{table:classify2}

\end{table}
}

\subsection{Classification with Semantic Analysis}

As described earlier, we have observed that online paid posters tend to post similar comments on the web, and based on this observation we have designed a simple method for semantic analysis. After integrating this semantic analysis method into our SVM model, we performed the test again over the Sohu dataset. The much improved performance results are shown in \emph{the last column} of Table~\ref{table:classify1}.

\nop{
\begin{table}[!hbp]

\centering

\begin{tabular}{|c|c|}
\hline
Metrics & 5-Feature Test \\\hline\hline

True Negative & 138 \\\hline

False Positive & 3 \\\hline

False Negative & 22 \\\hline

True Positive & 60 \\\hline

Precision & 95.24\% \\\hline 

Recall & 73.17\% \\\hline	

F-measure & 82.76\% \\\hline

Accuracy: & 88.79\% \\\hline

\end{tabular}

\caption{Test results with semantic analysis}

\label{table:classify3}

\end{table}
}

The results clearly demonstrate the benefit of using semantic analysis in the detection of online paid posters. The precision, recall, F-measure and accuracy have improved to $95.24\%$, $73.17\%$, $82.76\%$ and $88.79\%$, respectively. Based on these improved results, the semantic feature can be considered as a useful and important supplement to the other features. The reason why the semantic analysis improves performance is that online paid posters often try to post many comments with some minor edits on each post, leading to similar sentences. This helps the paid posters post many comments and complete their assignments quickly, but also helps our classifier to detect them.

\nop{This result not only proves that semantic analysis is very significant for the classification but also makes us conclude that such a SVM model is efficient to classify the online paid posters from the normal users.
}

\vspace{+0.1in}
\section{Related Work}\label{sec:relatedwork}
In this paper, we focus on paid posters who post comments online to influence people's thoughts regarding popular social events. We characterize the basic organization structure of paid posters as well as their online posting patterns. To the best of our knowledge, this paper is the first to study the social phenomenon of paid posters. 

Some of previous work regarding spam detection is similar to ours. Researchers have done plenty of work in this area to design better classification mechanisms. Niu {\em et al.}~\cite{9} conducted a quantitative study of forum spamming and found that forum spamming is a widespread problem and also developed a context-based detection method to identify spammers. Shin {\em et al.}~\cite{10} improve their work by designing a light-weight classifier that can be used on the forum server in real-time. They conducted detailed analysis on their datasets and identified typical features of forum spam to assist the classifier. Bhattarai {\em et al.}~\cite{11} explored the characteristics of comment spam in blogs based on their content. In order to detect the comment spam, they also investigated the notion of comment similarity through word duplication and semantic similarity.

The spammers in those scenarios use software to post malicious comments on their forums and blogs to change the results of search engine or to make theirs sites popular. However, the definition of spam has been extended to a much wider concept. Basically, any user whose behavior might interfere with normal communication or aid the spread of misleading information is specified as a spammer. Examples include forum spammers and comment spammers in social media. Yin {\em et al.}~\cite{12} studied so-called online harassment, in which a user intentionally annoys other users in a web community. They investigated the characteristics of a specific type of harassment using local features, sentimental features and contextual features. However, the performance of their detection model may still need to be improved since the maximum precision only reached $50\%$. Benevenuto {\em et al.}~\cite{13} proposed a detection mechanism to identify malicious users who post video response spam on Youtube. They studied the performance of several major attributes which were used to characterize the behavior of malicious users. Gao {\em et al.}~\cite{14} conducted a broad analysis on spam campaigns that occurred in Facebook network. From the dataset, they noticed that the majority of malicious accounts are compromised accounts, instead of ``fake" ones created for spamming. Such compromised accounts can be obtained through trading over a hidden online platform, according to \cite{15}.

\section{Conclusions and Future Work}\label{sec:conclusion}

Detection of paid posters behind social events is an interesting research topic and deserves further investigation. In this paper, we disclose the organizational structure of paid posters. We also collect real-world datasets that include abundant information about paid posters. We identify their special features and develop effective techniques to detect them. Our classifier based on SVM, with integrated semantic analysis, performs extremely well on the real-world case study. As future work, we plan to further improve our detection system and extend our research to other relevant areas, such as network marketing.

\nop{
We examine 5 different features of comments posters, including potential paid posters and normal users. Within these features, the percentage of replied comments and interval time distribution exhibit the capability to distinguish potential paid posters and normal users. Potential paid posters often don't click the "reply" button and will post comments in short time interval. We also list some hidden features of paid posters which are not appeared in the figure. Although we manually go through all the users' profiles and pick out 70 potential paid poster, the statistical analysis above doesn't make use of text mining which could be used to understand the meaning of comments. As a result, the performance of our classification still needs to be optimized. We consider it a future work.
}


\section*{Acknowledgment}
We thank Natural Sciences and Engineering Research Council of Canada (NSERC) for partial funding support and Mathematics of Information Technology And Complex Systems (MITACS) for supporting Xudong Zhang's internship at the University of Victoria. 
\balance
\bibliographystyle{abbrv}
\bibliography{introduction}

\end{document}